\documentstyle[11pt,newpasp,twoside,epsf]{article}
\newcommand{\mamo}[1]{\mbox{$#1$}}

\newcommand{\unit}[1]{\ifmmode \:\mbox{\rm #1}\else \mbox{#1}\fi}
\newcommand{\sbr}[1]{_{\rm #1}}

\newcommand{\mpc}{\unit{Mpc}}

\newcommand{\hmpcinvcub}{\mamo{h_{75}^{3} \mpc^{-3}}}

\def\edcomment#1{\iffalse\marginpar{\raggedright\sl#1\/}\else\relax\fi}
\reversemarginpar

\begin{document}
\title{Cosmology with the Space-Luminosity Distribution of Virialized Halos}
 \author{Christian Marinoni}
\affil{Department of Astronomy, Cornell University,\\
 412 Space Sciences Bldg. Ithaca, NY 14853, US.}
\author{Michael J. Hudson}
\affil{Department of Physics, University of Waterloo, \\ 
Waterloo, ON N2L 3G1, Canada.}
\author{Giuliano Giuricin}
\affil{Dipartimento di Astronomia, Universit\`a di Trieste, \\
via Tiepolo 11, 34131 Trieste, Italia.}

\begin{abstract}
We derive the observed
luminosity function of virialized systems. Coupling this statistic 
with  the mass function predicted by CDM cosmogonies 
we obtain the 
functional behavior of the mass--to--light ratio over a wide dynamical
range in mass.
We 
conclude that the mass--to--light ratio has a minimum close to the
luminosity of an $L_*$ galaxy halo. 
We 
also 
show how to derive in a self-consistent way the X-ray--to--optical
luminosity ratio for galaxy systems. As a consequence, we predict 
a fundamental break in the
auto-similar scaling behavior of fundamental correlations 
in going from the group to the cluster scales.

\end{abstract}

\section{Introduction}
The luminosity function 
of virialized systems (VS LF) is a relatively unexplored statistical tool.  

The VS LF can be used to probe 
a number of fundamental problems in cosmology, ranging from the
efficiency of galaxy formation to the connection between X-ray and
optical light in galaxy systems.  Moreover the VS LF is more robust at
the low--mass end of the clustering hierarchy than 
is the mass function obtained by 
traditional estimators such as projected velocity dispersions and
X-ray temperatures.

We constructed the VS LF (Marinoni, Hudson \& Giuricin, ApJ submitted)
after processing, with a weighting scheme, bound objects extracted from
the Nearby Optical Galaxy (NOG) catalog (Giuricin et al. 2000).  NOG
is a statistically controlled, distance-limited ($cz_{LG}\leq$ 6000
km/s) and magnitude-limited (B$\leq$14) complete sample of more than
$7000$ optical galaxies.  The sample covers 2/3 (8.27 sr) of the sky
($|b|>20^{\circ}$), a volume of $1.41 \times 10^{6}\hmpcinvcub$ and
has a redshift completeness of 98\% (see Marinoni 2001). 
The NOG sample spans a fair volume of the Universe, but better samples 
the nearby  cosmological space  than previous all-sky catalogues.

In Fig. 1, the NOG galaxy distribution is shown 
using 3D isodensity surfaces of low-density contrast ($\delta=1.5$)
obtained by smoothing with a Gaussian filter of short smoothing length
($r_s = 200$ km/s). In this way we show the richness of details recovered
also on small scales.

In Fig. 1, we have graphically summarized the performance
of the algorithms 
developed in order to reconstruct, in an self-consistent 
way, galaxy systems and pseudo real-space positions (Marinoni et
al. 1998; Marinoni et al. 1999; Giuricin et al. 2000).
Most of the NOG galaxies 
are found to be members of galaxy binaries (which comprise $\sim$15\%
of galaxies) or groups with at least three members ($\sim$45\% of
galaxies).  About 40\% of the galaxies are left ungrouped (isolated
galaxies). 

\section{The Virialized Systems Luminosity Function}

Having obtained the group catalog, we correct the sample for 
galaxies falling beyond the NOG magnitude limit and for the   
completeness of virialized systems of a given richness.      
The resulting ``all-systems'' catalog contains $\sim$ 4000 objects (
$\sim$ 2800 systems with one observed member and $\sim$1100 groups
with at least 2 members).
We then calculate a VS LF for virialized systems 
in the same manner as is usually done for galaxies.                     

Our results show that the B-band VS LF is insensitive to the choice of
the group-finding algorithm, to the peculiar velocity field models
with which distances are reconstructed, but is sensitive to the effects of the
large-scale environmental density.

We denote by the subscript ``s'' luminosities of systems (as opposed
to individual galaxies).  The VS LF (Fig. 2) is well described, over
the absolute magnitude range $-24.5 \leq M_s + 5 \log h_{75} \leq
-18.5$, by a Schechter function with $\alpha_s=-1.4 \pm 0.03$,
$M_s^{*} - 5 \log h_{75} =-23.1 \pm 0.06$ and $\phi_s^{*}=4.8 \times
10^{-4}\;\hmpcinvcub$ or by a double-power law: $\phi_{\rm pl}(L_s)
\propto L_s^{-1.45 \pm 0.07}$ for $L_s< L\sbr{pl}$ and $\phi(L_s)
\propto L_s^{-2.35 \pm 0.15}$ for $L_s > L\sbr{pl}$ with
$L\sbr{pl}=8.5 \times 10^{10} h_{75}^{-2} L_{\odot}$, corresponding to
$M_s- 5 \log h_{75} = -21.85$. The characteristic luminosity of
virialized systems, $L\sbr{pl}$, is $\sim 3$ times brighter than that
($L^{*}_{gal} = 2.7 \times 10^{10} h_{75}^{-2}L_{\odot}$) of the
luminosity function of NOG galaxies.  In Fig. 2, we show how isolated
galaxies dominate the low-luminosity portion of the VS LF, while richer
systems dominate in a progressive way the bright end.  In the limit
of low luminosity the VS LF approaches the galaxy LF.
We find that, after correcting for the unseen luminosity, systems with
total luminosity $L_s<L_{gal}^*$ give a 25\% contribution to the total
luminosity density $\rho_L$ of the universe, systems with $L_s< 5
L_{gal}^*$ and $L_s<10 L_{gal}^*$ contribute by 58\% and 75\%
respectively, while a substantial fraction of luminosity density
(10\%) is contributed by large systems with more than $30L_{gal}^*$.

\begin{figure}
\plotfiddle{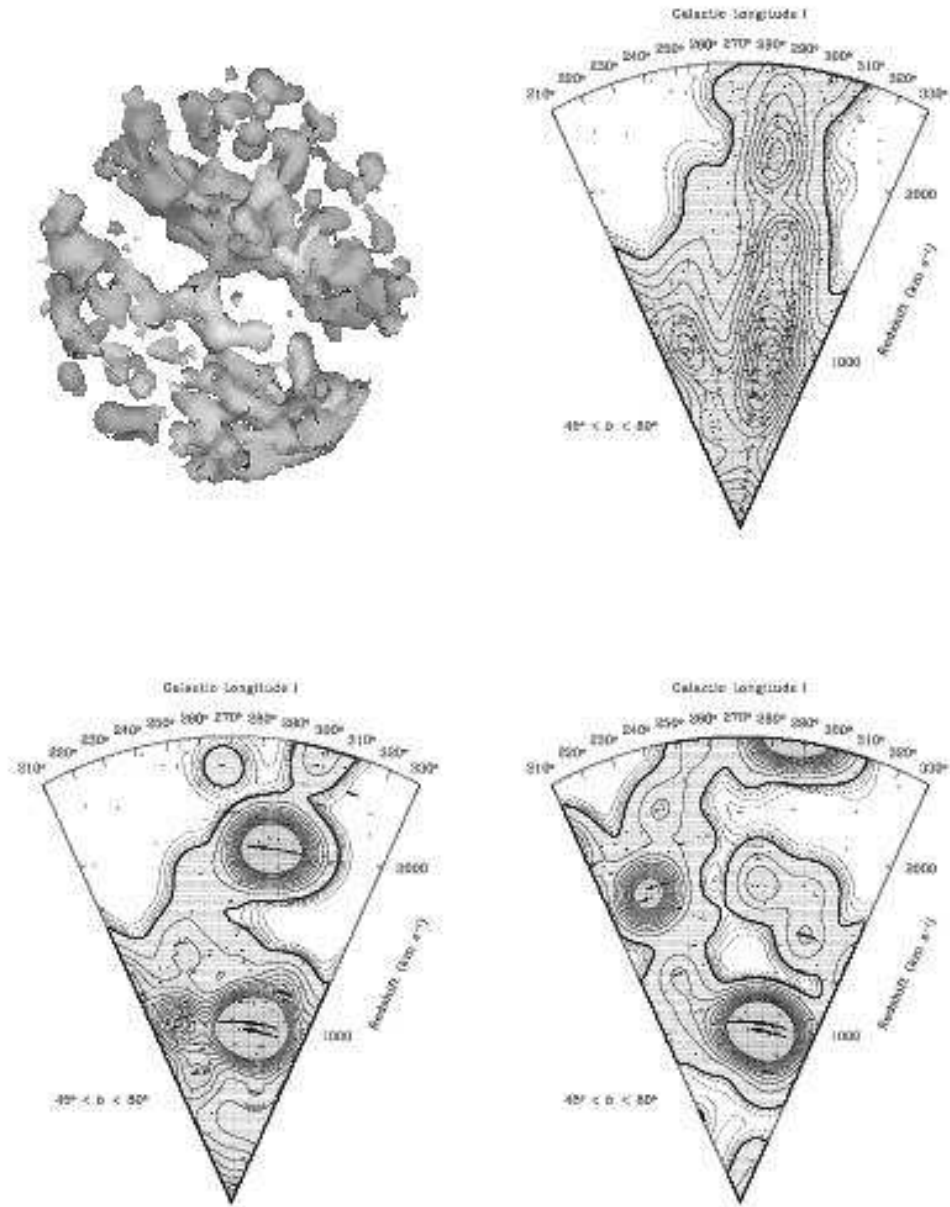}{13.5cm}{0}{93}{93}{-280}{-60}
\caption{{\em Upper left:} Density ($\delta$=1.5) distribution of NOG
galaxies smoothed using a Gaussian window function with smoothing
length $r=200$ km/s.  {\em Upper right:} redshift distribution of NOG
galaxies in a cone diagram centered on the Virgo region.  {\em Lower
left:} the same region after clustering reconstruction and after
having applied a particular peculiar velocity field model ({\em lower
right}).}
\end{figure}

\begin{figure}
\plotfiddle{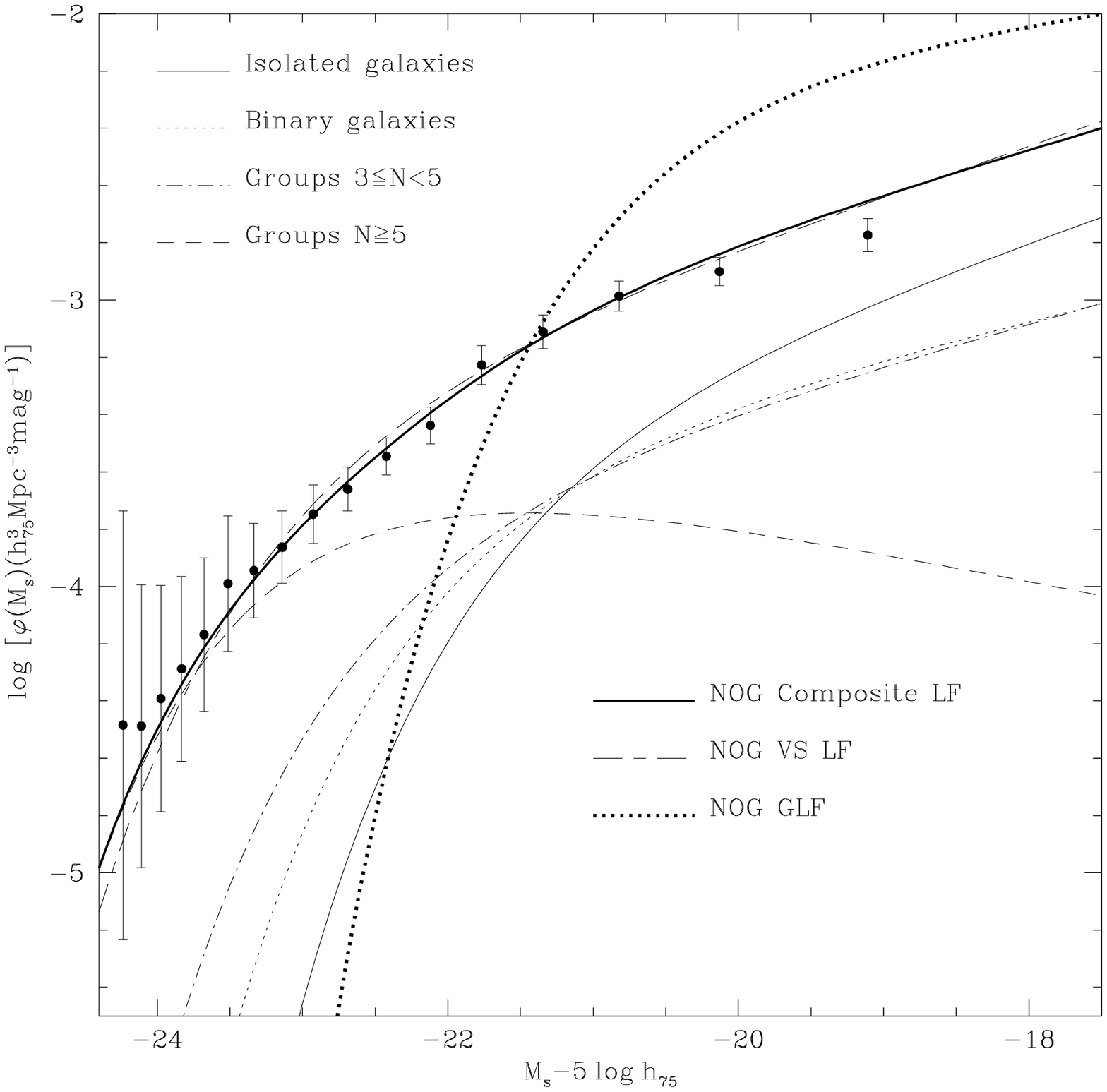}{6.5cm}{0}{40}{40}{-240}{-90}
\plotfiddle{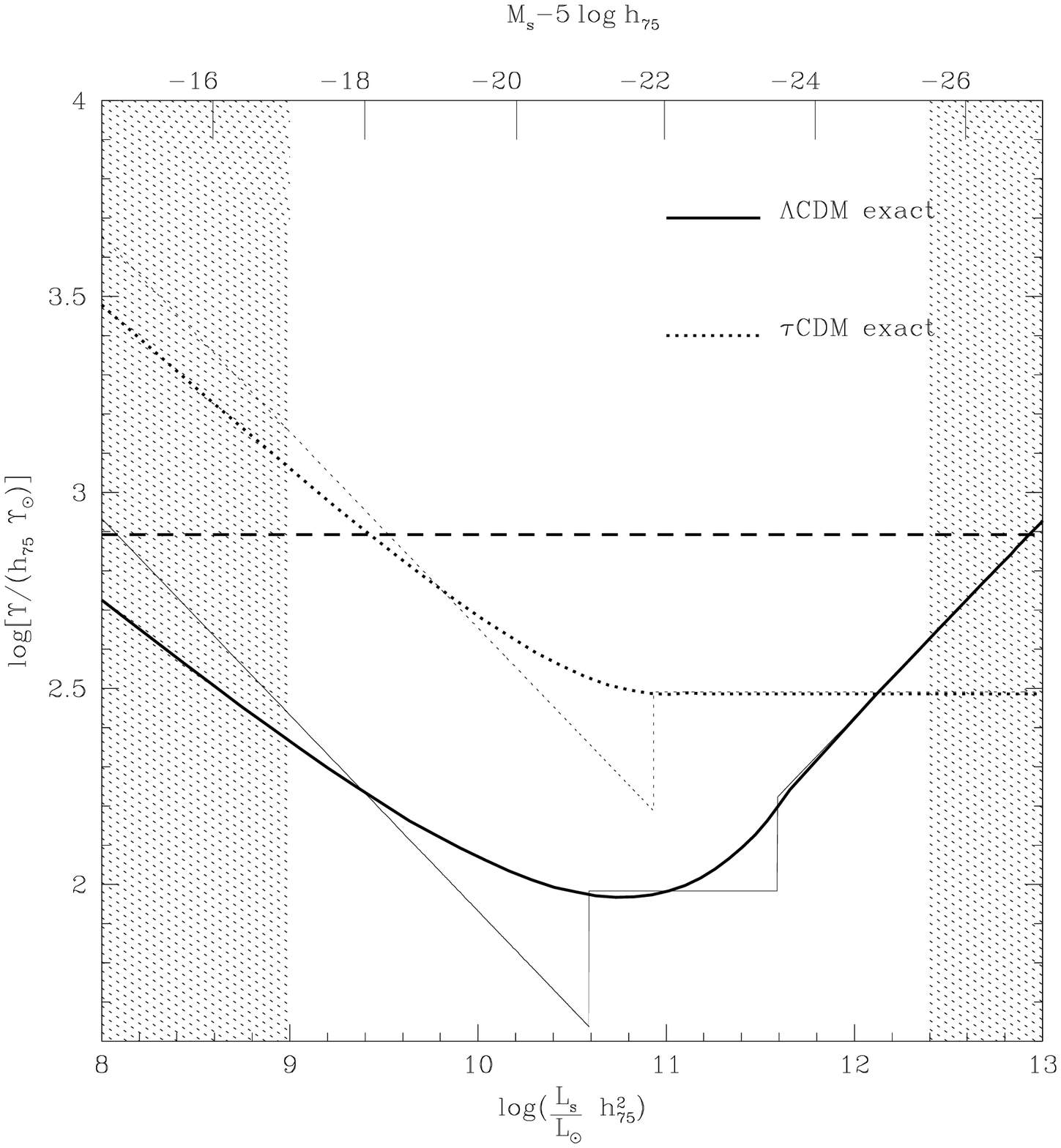}{0cm}{0}{40}{40}{0}{-60}
\caption{{\em Left:} we plot the LF of virialized haloes obtained by
summing the contribution from various systems. Points represent the
observed VS LF. The NOG galaxy LF (GLF) is shown for comparison. {\em
Right:} the Mass--to--light ratio ($\Upsilon$) is shown as a function
of the system absolute luminosity in the case of $\Lambda$CDM and
$\tau$CDM cosmogonies.}
\end{figure}

\section{Cosmology with Distribution Functions}

 The traditional approach for shaping interesting cosmological scaling
 relationships has been to estimate correlations directly from the
 data.  However this is problematic since different fitting procedures
 of the same set of data (with often unknown errors) usually lead to
 discrepant results.  Moreover, when errors are large it is almost
 impossible to disentangle subtle deviations from the traditional {\em
 null hyphotesis} of a logaritmically linear correlation.

 On the other hand, comparing statistical distributions of complete set
 of data is an economic way to obtain robust insights into the physical
 properties of cosmological objects. Here we show how comparing the VS
 LF with the Press \& Schechter mass distribution function and with the
 X-ray luminosity function we can link in a continuum some physical
 properties of galaxy and clusters.

\subsection{The Mass--to--Light Ratio as a Function of Environment}

The traditional assumption in cosmological applications is that
mass-to-light ratio ($\Upsilon$) is approximately constant.  
By comparing 
our observed luminosity function with the
Press-Schechter mass functions in cold dark matter cosmogonies (SCDM,
OCDM, $\tau$CDM and $\Lambda$CDM) 
we can turn the problem around and obtain the $\Upsilon$  function directly.

In this way the $\Upsilon$ scaling, which tells us about the efficiency with
which the universe transforms matter into light, can be used not only
as a traditional estimator of the density parameter $\Omega_0$ but
also as a diagnostic tool for different 
models of galaxy formation. 

We find that, if a constant $\Upsilon$ is
assumed, all cosmological models fail to match our results.  In order
for these models to match the faint end of the luminosity function, a
$\Upsilon$ decreasing as $L^{-0.5 \pm 0.06}$ with luminosity
is required.  The value of $\Upsilon$ reaches a minimum around
$10^{10.5}\;h^{-2}_{75}\;L_{\odot}$, which corresponds roughly to
$L_{gal}^*$.  It then remains quite flat over 1.5 dex in luminosity.
The behavior at the bright end is model-dependent; for the $\tau$CDM
model, $\Upsilon$ must remain constant ($m \propto L^{1\pm 0.1}$),
whereas for the $\Lambda$CDM model it must increase with luminosity as
$L^{0.50 \pm 0.26)}$ from the scale of galaxy groups ($\sim 
10^{13} h_{75}^{-1}\; m_{\odot}$) to that of rich clusters ($ \sim 
10^{15} h_{75}^{-1}\; m_{\odot}$).  This scaling behavior of $\Upsilon$ 
in a $\Lambda$CDM cosmology appears to be in
qualitative agreement with the predictions of semi-analytical models
of galaxy formation (Benson et al 2000).

\begin{figure}
\plotfiddle{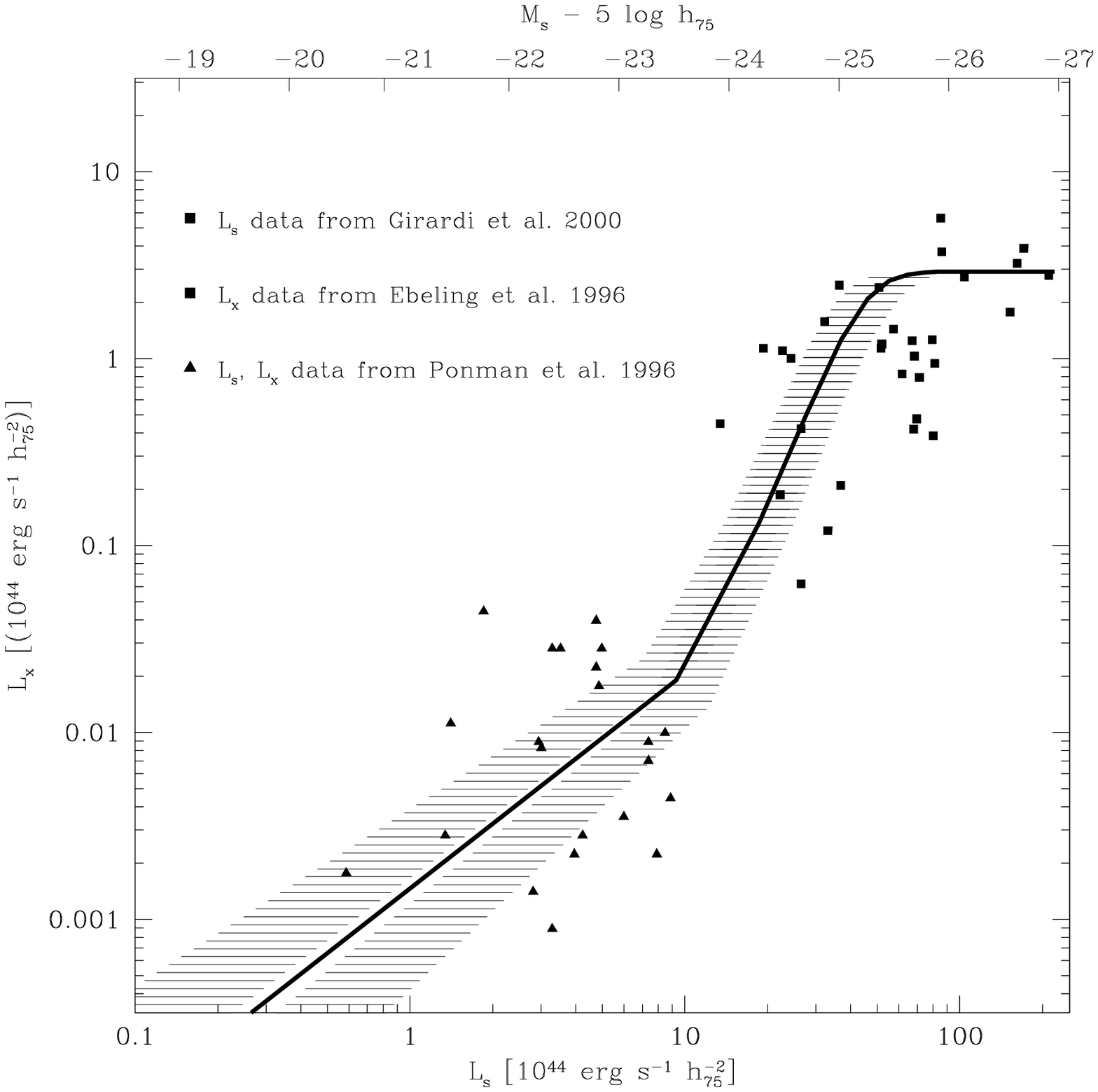}{6.5cm}{0}{43}{43}{-240}{-110}
\plotfiddle{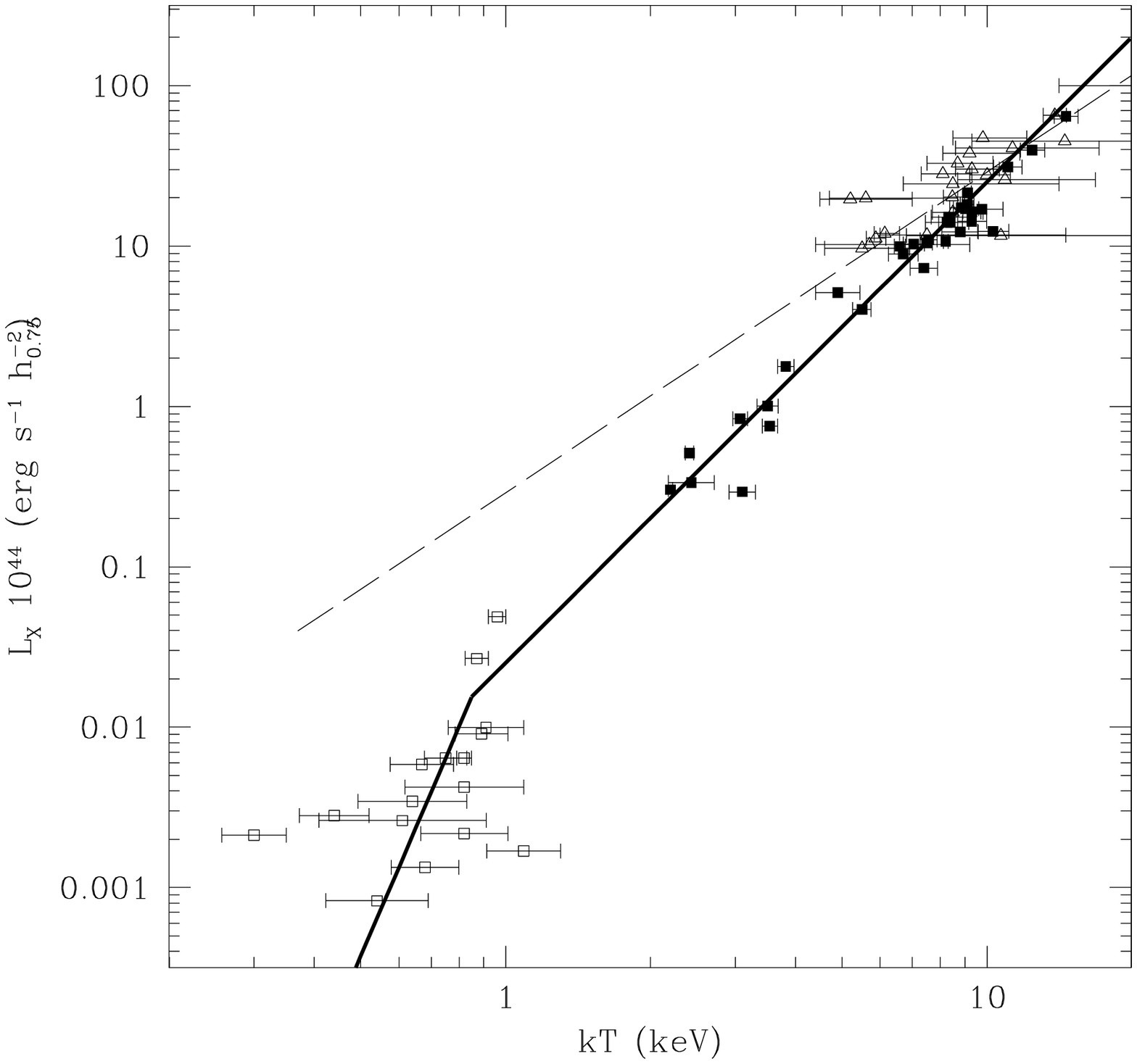}{0cm}{0}{43}{44}{0}{-90}
\caption{{\em Left:} the relation between optical ($L_s$) and X-ray
($L_x$) luminosities is shown together with the observed luminosities
of groups and clusters.  The hatched region corresponds to the errors
in the $L_x-L_s$ relation. {\em Right:} the relation between X-ray
luminosity and temperature in a $\Lambda$CDM cosmology.  The dashed
line refers to the self-similar model of Kaiser (1986).
}
\end{figure}

\subsection{The $L_X$ vs $L_s$ Function}

The problem of deriving the functional form $L_x=f(L_s)$ in galaxy systems
can be solved  if we assume 
that all virialized halos detected in the optical also radiate in 
X-rays and compare their observed LFs.  In the regime of faint
systems and groups ($M_s > -23 +5\, \log \, h_{75}$ or $L_x <
10^{42}h_{75}^{-2} {\rm erg}\;{\rm s}^{-1}$), 
we find that 
the numerical solution is well
approximated by $L_X \propto L_s^{1.15}$.  In the regime of rich groups
to clusters ($-25 < M_s- 5\, \log \, h_{75} < -23$ or $ 10^{42} < L_x
(h_{75}^{-2}{\rm erg}\;{\rm s}^{-1}) < 5 \cdot 10^{44}$), the solution is
well approximated by $L_x\propto L_s^{3.5} $.  In Fig. 4 we compare
our predictions with observational data. Interestingly we note that
the change in the shape of the $L_x-L_s$ relation for $L_x <
10^{42} h_{75}^{-1} {\rm erg} \; {\rm s}^{-1}$ 
corresponds exactly to a 
change in the chemical properties and the spatial distribution of the
ICM on the scales of groups (Renzini 1997).

\begin{table}[h]    
\footnotesize{
\caption{Scaling-Law Matrix}
\begin{tabular}{||l|l|l|l||}

\hline
\hline
\multicolumn{2}{||c|}{{\large {\bf Groups}}} &  \multicolumn{2}{|c||}{{\large 
{\bf Clusters}}}  \\

\hline
{\bf Predicted} & {\bf Observed} & {\bf Predicted} & {\bf Observed}  \\
\hline
\hline
$\mathbf{L_x\propto L^{1.5 \pm 0.3}}        $ &  & $\mathbf{L_x \propto L^{3.5\pm 0.7}}$&  \\
 & & & \\
\hline
$\mathbf{L\propto \sigma^{3\pm 0.45}}$& $L\propto \sigma^{3}$  & $\mathbf{L\propto\sigma^{1.7\pm 0.38}}$&$\mathbf{L\propto \sigma^{1.56}}$\\
&Moore et al. 1993 & & Adami et al. 1998 \\
\hline

$\mathbf{L_x \propto \sigma^{3.9\pm 1.2}}$&$\mathbf{L_x \propto \sigma^{5\pm 2.1}}$& $\mathbf{L_x\propto \sigma^{6 \pm 1.3}}$&$\mathbf{L_x\propto\sigma^{6.38 \pm 0.46}}$\\
&Ponman et al. 1996 & & White, et al. 1997 \\
& $\mathbf{L_x\propto\sigma^{5 \pm 0.9}}$ & & \\
&Helsdon et al. 2000 & &  \\
\hline

$\mathbf{L_x \propto T^{3.9 \pm 1.2}}$&
$\mathbf{L_x \propto T^{8.2 \pm 2.1}}$& 
$\mathbf{L_x\propto T^{3.0 \pm 0.65}}$&
$\mathbf{L_x\propto T^{3.16 \pm 0.11}}$\\
& Ponman et al 1996 & &White et al. 1997\\
\hline

$\mathbf{\frac{n}{\rho_*}\propto M^{2/3}}$ & &$\mathbf{\frac{n}{\rho_*}\propto M^{1/3}}$ & \\
 & & & \\

\hline
\hline
\end{tabular}
}
\end{table}

Coupling the mass--to--light ratio with the X-ray--to--optical
luminosity ratio and making some minimal and 
self-consistent 
hypotheses (virialization, X-ray emission described in terms of
thermal bremsstrahlung), we can predict the behavior of fundamental
scaling relationships over the group and cluster regimes.  In Table 1,
we compare our predictions ($\Lambda$CDM cosmology) with existing
observations. Our results are in agreement with models that predict an
initial non-equilibrium phase of preheating during halo formation
(Cavaliere, Menci, \& Tozzi 1997).

\end{document}